\documentstyle[prb,aps,epsfig,amsmath,amssymb,multicol]{revtex}

\begin{document}

\title{2D scaling approach for the first order transition in ${\bf %
Bi_2Sr_2CaCu_2O_{8+y}}$}
\author{T. Schneider and J. M. Singer}
\address{Physik-Institut, Universit\"at Z\"urich, Winterthurerstr. 190, CH-8057
Z\"urich, Switzerland}
\date{\today}
\maketitle

\begin{abstract}
We analyze recent measurements by Ooi et al.
of the angular dependence of the first order
vortex phase transition in $\mathrm{Bi_2Sr_2CaCu_2O_{8+y}}$.
The experimental findings of these authors are interpreted
in terms of a $2D$ versus $3D$ scaling approach. 
Invoking a crossover from a 3D to a 2D scaling approach, it shown, that
the apparent failure of a $3D$ scaling approach away from the bulk transition
temperature is due to a crossover to decoupled independent superconducting
slabs. 
\end{abstract}

\pacs{74.25.-q,74.60.Ec,74.25.Dw,74.25.Ha}

\noindent PACS: 74.25.-q, 74.60.Ec, 74.25.Dw, 74.25.Ha

\bigskip

\begin{multicols}{2}

In a recent letter, Ooi et al. \cite{Ooi} investigated the angular
dependence of the first order transition (FOT) field, $H_{FOT}\left( \delta
\right) $) in the mixed state of  ${\rm Bi_{2}Sr_{2}CaCu_{2}O_{8+y}}$
(BSCCO) by means of local magnetization measurements. Here, $\delta $  is
the angle between the applied magnetic field and the c-axis.The FOT has been
detected as a step in the magnetization \cite{Zeldov}. The authors report
results for four samples: optimally doped, slightly overdoped, overdoped,
and highly overdoped. The critical temperatures are $T_{c}\left( H=0\right)
=86.7,\ 85.5,\ 80.3$ and $76.8\ K$, respectively. The reported measurements
have been performed considerably below $T_{c}\left( H=0\right) $ (e.g., at $%
T=75K$ for the  slightly overdoped sample with $T_{c}\left( H=0\right) =85.5K
$ . They conclude that the experimental data  for $H_{FOT}\left( \delta
\right) $, taken at fixed temperature, is inconsistent with the scaling law
resulting from the  $3D$ anisotropic Ginzburg-Landau theory  and they
propose a `new' scaling law for the angular dependence of the vortex phase
transitions, which appears to be consistent with the experimental data.

As Ooi et al. state, this failure points to a large effective mass
anisotropy. In highly anisotropic superconductors such as BSCCO and,
similarly, in heavily underdoped materials, where the effective mass
anisotropy $\gamma $ increases markedly by approaching the underdoped limit,
quasi-$2D$ behavior is expected to occur as long as 
\begin{equation}
\xi _{\perp }(T)=\xi _{\perp ,0}|t|^{-\nu }\leq s,\quad t=\frac{T-T_{c}}{%
T_{c}},\quad \nu =2/3,
\end{equation}
where $\xi _{\perp }$ is the correlation length along the crystallographic
c-axis, and $\xi _{\perp ,0}$ is the corresponding critical amplitude. $s$
denotes the spatial interlayer separation, and $T_{c}$ is the bulk
transition temperature. For $\gamma $ sufficiently large $\xi _{\perp ,0}\ll
s$. Before the interlayer coupling causes $3D$ order, $\xi _{\perp }$ has to
grow larger than the spacing $s$ between the slabs of thickness $d_{s}$.
Thus, $3D$ fluctuation effects can be expected only for 
\begin{equation}
|t|\ll t_{cross}=\left( \frac{\xi _{\perp ,0}}{s}\right) ^{1/\nu }=\left( 
\frac{\xi _{\parallel ,0}}{\gamma s}\right) ^{1/\nu },\quad \nu \approx 2/3,
\end{equation}
while for $t\geq t_{cross}$ $2D$ fluctuations are relevant. In materials
such as BSCCO, exhibiting a pronounced effective mass anisotropy, $t_{cross}$
turns out to be rather small and the correlation length perpendicular to the
layers can become much smaller than the slab separation $s$. In the
corresponding temperature regime, these materials mimic to a stack of nearly
independent superonducting slabs of thickness $d_{s}$. This modelling is
also well confirmed by, e.g., the so-called crossing point phenomenon \cite
{TSJMS_physica1999}. For a sheet of thickness $d_{s}$ the singular part of
the free energy is expected to adopt the scaling form \cite
{TSJMS_physica1999,BOOK} 
\begin{equation}
f_{s}=\frac{k_{B}TQ_{2}^{\pm }}{(\xi _{\Vert }^{\pm })^{2}d_{S}}G_{2}^{\pm }(%
{\cal Z}),
\end{equation}
where $G_{2}$ is an universal $2D$ scaling function of its argument ${\cal Z}
$ \cite{TSJMS_physica1999,SchneiderTanner,BOOK}, 
\begin{equation}
{\cal Z}=\left( H(\xi _{\parallel }^{\pm })^{2}|\cos (\delta )|/\Phi
_{0}+H^{2}(\xi _{\Vert }^{\pm })^{2}d_{S}^{2}\sin ^{2}(\delta )/\Phi
_{0}^{2}\right) ^{1/2}.
\end{equation}
supposed there is a melting transition of the vortex lattice, the universal
scaling function $G_{2}^{\pm }({\cal Z})$ will exhibit a singularity at some
fixed value ${\cal Z}={\cal Z}_{m}$. Accordingly, the angular dependence of
the melting field, where for fixed $T$ the transition occurs, is then given
by 
\begin{equation}
\widetilde{H}_{s}(\delta )=\frac{|\cos (\delta )|}{2a\sin ^{2}(\delta )}%
\left( -1+\sqrt{1+4a^{2}\tan ^{2}(\delta )}\right) ,  \label{ooi2D}
\end{equation}
where 
\begin{equation}
\widetilde{H}_{s}(\delta )=\frac{H_{m}(\delta )}{H_{m}(\delta =0)},\quad a=%
\frac{{\cal Z}_{m}^{2}d_{s}^{2}}{\xi _{\Vert }^{2}}.
\end{equation}
However, as long as the correlation length $\xi _{\bot }$ is much larger
than the separation $s$ of the superconducting sheets, the expression for
anisotropic bulk systems should apply. Here, the scaling variable is given
by \cite{Ooi,SchneiderTanner}
\begin{equation}
{\cal Z}=\frac{\xi _{\Vert }^{2}H}{\Phi _{0}}\left( \cos ^{2}(\delta )+\frac{%
1}{\gamma ^{2}}\sin ^{2}(\delta )\right) ^{1/2},
\end{equation}
so that 
\begin{equation}
\widetilde{H}_{b}(\delta )=\left( \cos ^{2}(\delta )+\frac{1}{\gamma ^{2}}%
\sin ^{2}(\delta )\right) ^{-1/2}.  \label{ooi3D}
\end{equation}

{\narrowtext
\begin{figure}
\centerline{\epsfig{file=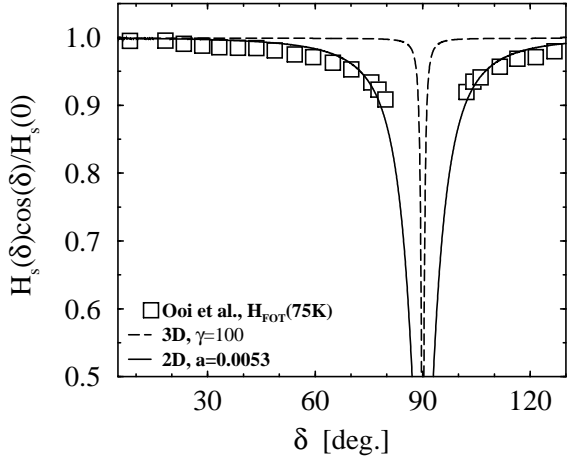,width=8cm,angle=0}}
\caption{Angular dependence $H(\protect\delta )\cos (\protect\delta )/H(0)$
of the FOT field, $T=75K$, as measured by Ooi et al.$^1$ for a slightly
overdoped BSCCO crystal with $T_{c}=85.5K$ ($\square $). The solid line is a
fit to Eq. (\ref{ooi2D}), whereas the dashed line uses Eq. (\ref{ooi3D}). 
\label{fig1}}
\end{figure}}

The resulting angular dependences of $\widetilde{H}$ for the $2D$ and $3D$
approach, as well as  measured data of Ooi et al  are depicted in Fig. \ref
{fig1}. The experimental data shown in this Fig. \ref{fig1} clearly reveal
the failure of $3D$ approach for this temperature regime and point to the
applicability Eq. (\ref{ooi2D}). A fit, yielding $a\approx 0.0053$ for
$T=75K$, leads to a rather reasonable agreement.
Noting that $\xi_\parallel$ decreases as the temperature is lowered
($T<T_c$) and with that $a$ increases it also follows that the width of
$H(\delta)$ around $\delta\approx 90^o$ becomes larger. This behavior
is fully consistent with the experimental results of Ooi et al. \cite{Ooi}.

Even though these data uncover the $2D$ behavior, they do not provide
evidence for the $3D$ to $2D$ crossover. This crossover has been studied,
both experimentally \cite{Silva,Tanner} and theoretically \cite
{SchneiderTanner,Tanner} in terms of the angular dependence of the onset
field $H^{\ast }$, where a measurable resistance is observed. According to
the dynamic scaling approach \cite{BOOK} this field is expected to adopt the
scaling forms given by Eqs. (\ref{ooi2D}) and (\ref{ooi3D}).

Noting that Eq. (\ref{ooi2D}) leads for $\delta $ close to $90^{o}$ to a
cusplike structure, while Eq. (\ref{ooi3D}) corresponds to a bellshaped
behavior, measurements with high angular resolution around $\delta =90^{o}$
and performed below $T_{c}\left( H=0\right) =84.1K$ \ should exhibit the
crossover between these limiting cases. This expectation is nicely confirmed
by the measured angular dependence of $H^{\ast }$ in films of the same BSCCO
compound shown in Fig. \ref{fig2}. Apparently at $T=79.5K$ (i.e., $%
t=5.55\cdot 10^{-2}$) the data are fully consistent with $2D$ behavior,
yielding $a\approx 0.0050$ and $H^{\ast }(\delta =0)=0.984T$, while at $%
T=82.8K$ ($t=1.55\cdot 10^{-2}$) $3D$ behavior dominates with $\gamma
\approx 77$, $H^{\ast }(\delta =0)=9.513\cdot 10^{-4}T$.

{\narrowtext
\begin{figure}
\centerline{\epsfig{file=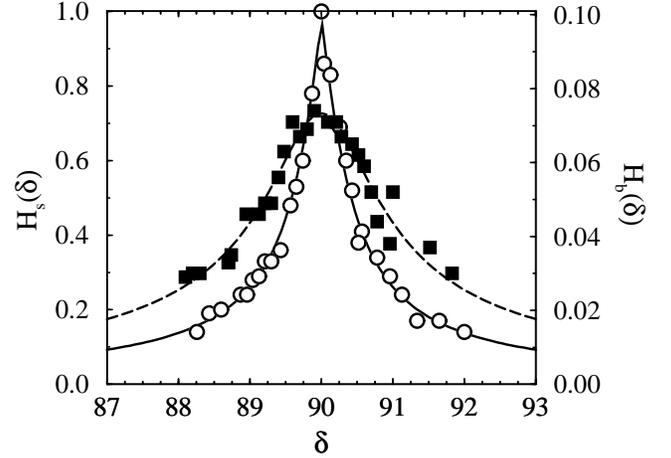,width=8cm,angle=0}}
\caption{Angular dependence of the resistence onset field $H^{\ast }(\protect%
\delta )$ for BSCCO with $T_{c}\left( H=0\right) =84.1K$, data taken from
Tanner et al. $^7$, $T=82.8K$ ($\blacksquare $) and $T=79.5K$ ($\circ $). The
solid line corresponds to Eq. (\ref{ooi2D}), yielding $a\approx 0.0050$,
whereas the dashed line denotes a fit to Eq. (\ref{ooi3D}) with $\protect%
\gamma \approx 77$.
\label{fig2}}
\end{figure}}

In conclusion, the behavior in the intermediate temperature regime mirrors
the dimensional crossover between the limiting $2D$-$XY$ and $3D$-$XY$
behavior. Our analysis clearly reveals that the continuum description only
holds rather close to $T_{c}$. In particular, estimates of the anisotropy $%
\gamma $ must be deduced close to criticality in order to avoid unphysical
fitting to a mere crossover regime. Finally, there is no need for a
fundamentally different scaling approach as suggested by Ooi et al. \cite{Ooi}.

\end{multicols}

\end{document}